\documentclass{article}

\PassOptionsToPackage{numbers,sort&compress}{natbib}
\usepackage[final]{neurips_2024}

\usepackage[utf8]{inputenc}   
\usepackage[T1]{fontenc}      
\usepackage{hyperref}         
\usepackage{url}              
\usepackage{booktabs}         
\usepackage{amsfonts}         
\usepackage{amsmath}          
\usepackage{nicefrac}         
\usepackage{microtype}        
\usepackage{xcolor}           
\usepackage{graphicx}         
\usepackage{subcaption}       
\usepackage{multirow}         
\usepackage{cleveref}         
\usepackage{enumitem}         
\usepackage{amssymb}

\title{MEG-Mamba: A Scalable State-Space Foundation Model for Magnetoencephalography}

\author{%
Chetan Gohil\thanks{Corresponding author: \texttt{chetan.gohil@ndcn.ox.ac.uk}.} \\
OHBA, Department of Psychiatry \\
University of Oxford \\
\And
SungJun Cho\\
Nuffield Department of Clinical Neurosciences \\
University of Oxford \\
\And
Oiwi Parker Jones\\
Department of Engineering Science \\
University of Oxford \\
\And
Mark Woolrich\\
OHBA, Department of Psychiatry \\
University of Oxford \\
}

\begin{document}

\maketitle

\begin{abstract}
Magnetoencephalography (MEG) is an imaging technique that offers a non-invasive, millisecond-resolution view of human brain activity. The increasing availability of MEG data presents an opportunity to take advantage of a recent advance in artificial intelligence, namely self-supervised foundation models. Existing foundation models for MEG (and electroencephalography) have been built on a transformer architecture. Here, we introduce MEG-Mamba: a generative foundation model for neural activity (source reconstructed, parcellated MEG) built on a Mamba architecture. MEG-Mamba is trained to predict the next token of a discretised MEG signal, conditioned on a brain region and recording session. MEG-Mamba surpasses the generative fidelity of a transformer-based alternative (MEG-GPT) while pre-training in less time (22 vs 400 GPU-hours) and modelling a longer context (4 vs 0.32\,s). We evaluate MEG-Mamba by examining the spatio-spectral characteristics of the data it generates and by interpreting its learned embeddings. Furthermore, we demonstrate that lightweight stimulus conditioning (LoRA) can be used to generate realistic task-evoked responses that are not included in pre-training. Our results suggest that Mamba is a promising backbone for scaling neural foundation models.
\end{abstract}

\section{Introduction} \label{sec:introduction}

Magnetoencephalography (MEG) is a brain imaging technique that measures the magnetic fields produced by neuronal populations~\citep{hamalainen1993meg,baillet2017meg}. MEG uniquely captures fast (millisecond) neuronal dynamics, which makes it a powerful tool for both basic neuroscience and applied settings, such as brain-computer interfaces and the assessment of neurological and psychiatric disorders.

MEG datasets are often collected for specific cohorts or during specific cognitive tasks. Consequently, datasets are typically small ($\lesssim$100 participants), limiting the reach of modern artificial intelligence methods. Foundation models that pool data across a large population and different tasks offer a potential solution~\citep{bommasani2021foundation}. Such models have transformed natural language~\citep{devlin2019bert,brown2020gpt3} and computer vision~\citep{dosovitskiy2021vit,he2022mae}.

Most neural foundation models have targeted electroencephalography (EEG; \Cref{tab:fms}), which is inexpensive and widely available~\citep{jiang2024labram,wang2025cbramod,elouahidi2025reve}. Foundation models for MEG are now emerging~\citep{huang2026meggpt,jayalath2026megxl,xiao2025brainomni,tholke2026roadmap}. MEG has the advantage that its underlying signal (the magnetic field from neuronal populations) passes straight through the head. Therefore, MEG supports more accurate \emph{source localisation}~\citep{vanveen1997beamformer,baillet2017meg}, yielding a neural representation that is spatially interpretable and directly comparable across subjects and scanners.

Existing MEG foundation models (e.g. MEG-GPT~\citep{huang2026meggpt}) have been built with transformer backbones whose self-attention scales quadratically with sequence length. This forces short context windows and results in a high training cost. State-space models offer an alternative: the Mamba family~\citep{gu2023mamba,dao2024mamba2,lahoti2026mamba3} replaces attention with an input-dependent recurrence that scales linearly with sequence length, enabling much longer contexts at far lower cost. The most recent of these, Mamba-3~\citep{lahoti2026mamba3}, is a natural fit for MEG: its complex-valued state space can represent oscillatory modes~\citep{gu2022s4d,rusch2025linoss}, which are a prominent characteristic of neuronal activity~\citep{buzsaki2004oscillations, ward2003oscillations}.

Our contributions are as follows:
\begin{itemize}[leftmargin=*]
\item \textbf{MEG-Mamba.} The first autoregressive foundation model for MEG built on a Mamba-3 backbone.
\item \textbf{Efficiency.} Compared to MEG-GPT on the same dataset, MEG-Mamba pre-trains in $22$ vs $400$ GPU-hours and models a 4 vs 0.32\,s context.
\item \textbf{Fidelity.} MEG-Mamba surpasses the generative fidelity of MEG-GPT, reproducing the spatial, spectral, and temporal structure of real MEG data.
\item \textbf{Zero-shot generalisation.} MEG-Mamba does not use a learned lookup table for session embeddings (as is done in MEG-GPT). This allows us to apply MEG-Mamba to new subjects without fine-tuning.
\item \textbf{Controllable generation.} We show that, although pre-trained on resting-state data, the model can be steered by a lightweight adapter (a learned stimulus embedding plus LoRA~\citep{hu2022lora}) to generate realistic task-evoked responses.
\end{itemize}

\section{Background and Related Work} \label{sec:related}

\paragraph{Foundation models for brain electrophysiology.} Self-supervised foundation models for neural data have advanced rapidly, the majority targeting EEG and intracranial EEG (iEEG), where large corpora are available~\citep{kuruppu2025critical}. Pre-training datasets have grown from thousands of hours~\citep{jiang2024labram,wang2025cbramod,zhou2025csbrain} to tens of thousands~\citep{yuan2024brant2,elouahidi2025reve,chen2026unintfm}, and models to over a billion parameters~\citep{yue2024braingpt,jiang2025neurolm,han2025diver1}. We focus on \emph{generative} models, trained to predict the signal itself, and group them by pre-training objective: masked reconstruction, which infills held-out portions of the signal~\citep{wang2023brainbert,zhang2023brant,jiang2024labram,xiao2025brainomni,wang2024eegpt,wang2025cbramod,yuan2024brant2,elouahidi2025reve,zhou2025csbrain}, and autoregressive next-step prediction, which generates the signal one step at a time~\citep{cui2024neurogpt,yue2024braingpt,jiang2025neurolm,cui2026brainrvq}. Of these, the autoregressive models define a proper distribution that can be sampled to generate complete recordings. Flow matching has recently been explored as a continuous generative alternative~\citep{hwang2026bfm2}. We omit purely contrastive and latent-prediction models, such as BENDR~\citep{kostas2021bendr}, BIOT~\citep{yang2023biot}, and JEPA-style approaches~\citep{panchavati2026laya}, which learn representations without a generative objective.

\begin{table}[!ht]
  \caption{\textbf{Generative neural foundation models.} ``Backbone'' is Transformer (Tr.), state-space model (SSM), or U-Net. ``Obj.'' is the objective: autoregressive (AR), masked reconstruction (MR), or flow matching (FM). ``Data'' is the pre-training corpus duration in hours; ``Ch.'' the number of input channels (parcels for source-space models); ``Rate'' the sampling rate of the model input; ``Params'' the largest pre-trained variant reported; ``Compute'' the reported pre-training cost in GPU-hours (GPU types differ between papers). ``--'' denotes not reported; $\dagger$ marks papers that report the GPU type and count but not the training time.}
  \label{tab:fms}
  \centering
  \footnotesize
  \setlength{\tabcolsep}{3.5pt}
  \resizebox{\linewidth}{!}{%
  \begin{tabular}{llllrlrrl}
    \toprule
    Model & Modality & Backbone & Obj. & Data\,(h) & Ch. & Rate\,(Hz) & Params & Compute\,(GPU-h) \\
    \midrule
    \textbf{MEG-Mamba} & \textbf{MEG} & \textbf{SSM} & \textbf{AR} & \textbf{87} & \textbf{100} & \textbf{250} & \textbf{3.4M} & \textbf{22} \\
    MEG-GPT~\citep{huang2026meggpt} & MEG & Tr. & AR & 97 & 52 & 250 & 7.4M & ${\sim}400$ \\
    MEG-XL~\citep{jayalath2026megxl} & MEG & Tr. & MR & ${\sim}300$ & var. & 50 & 20M & ${\sim}12$ \\
    BrainOmni~\citep{xiao2025brainomni} & EEG+MEG & Tr. & MR & 2{,}653 & 19--306 & 256 & 33M & ${\sim}290$ \\
    Brain-OF~\citep{guo2026brainof} & fMRI+EEG+MEG & Tr. & MR & -- & var. & 200 & 1.7B & ${\leq}32{,}000$ \\
    \midrule
    BrainBERT~\citep{wang2023brainbert} & iEEG & Tr. & MR & 44 & 1{,}249 & 2{,}000 & -- & -- \\
    Brant~\citep{zhang2023brant} & iEEG & Tr. & MR & 2{,}528 & 52--153 & 250 & 506M & ${\sim}270$ \\
    BrainWave (Brant-2)~\citep{yuan2024brant2} & iEEG+EEG & Tr. & MR & 40{,}907 & var. & var. & -- & ${\sim}400$ \\
    DIVER-1~\citep{han2025diver1} & iEEG & Tr. & MR & 5{,}310 & var. & 500 & 1.8B & --$^\dagger$ \\
    \midrule
    LaBraM~\citep{jiang2024labram} & EEG & Tr. & MR & 2{,}535 & var. & 200 & 369M & --$^\dagger$ \\
    Neuro-GPT~\citep{cui2024neurogpt} & EEG & Tr. & AR & 5{,}656 & 22 & 250 & -- & -- \\
    EEGPT~\citep{wang2024eegpt} & EEG & Tr. & MR & -- & 58 & 256 & 101M & --$^\dagger$ \\
    BrainGPT~\citep{yue2024braingpt} & EEG & Tr. & AR & -- & $\leq$138 & 256 & 1.1B & --$^\dagger$ \\
    CBraMod~\citep{wang2025cbramod} & EEG & Tr. & MR & ${>}9{,}000$ & 19 & 200 & 4M & ${\sim}480$ \\
    NeuroLM~\citep{jiang2025neurolm} & EEG & Tr. & AR & ${\sim}25{,}000$ & var. & 200 & 1.7B & --$^\dagger$ \\
    FEMBA~\citep{tegon2025femba} & EEG & SSM & MR & ${>}21{,}000$ & 22 & 256 & 386M & -- \\
    CSBrain~\citep{zhou2025csbrain} & EEG & Tr. & MR & ${>}9{,}000$ & 19 & 200 & -- & ${\sim}400$ \\
    REVE~\citep{elouahidi2025reve} & EEG & Tr. & MR & 61{,}415 & var. & 200 & 408M & ${\sim}260$ (est.) \\
    BrainRVQ~\citep{cui2026brainrvq} & EEG & Tr. & AR/MR & ${\sim}9{,}200$ & 19 & 200 & 5.8M & ${\sim}120$ \\
    B[FM]$^2$~\citep{hwang2026bfm2} & EEG & U-Net & FM & ${\sim}307$ & 19 & 200 & 6M & --$^\dagger$ \\
    \bottomrule
  \end{tabular}}
\end{table}

Among MEG foundation models, MEG-GPT~\citep{huang2026meggpt} is the closest to ours, pre-training a transformer with next-step prediction on tokenised MEG data; MEG-XL~\citep{jayalath2026megxl} uses masked prediction with a minutes-long context for brain-to-text decoding; and BrainOmni~\citep{xiao2025brainomni} and Brain-OF~\citep{guo2026brainof} model EEG and MEG jointly (Brain-OF also fMRI) with masked objectives. Clinical~\citep{liao2026headache} and speech-decoding~\citep{jayalath2025bitter} applications have also been reported. All of these MEG models are built on transformer backbones. \Cref{tab:fms} situates MEG-Mamba among generative neural foundation models. Reported pre-training compute is rare; where it is available, MEG-Mamba is among the cheapest models to pre-train.

\paragraph{Scaling and evaluating neural foundation models.} Recent benchmarks suggest the current scale of models and data does not deliver transferable representations: across many downstream tasks, large EEG foundation models have not reliably outperformed specialist models trained from scratch, or even classical features~\citep{xiong2026eegfmbench,kastrati2025eegbench,liu2026eegfmcompass,shen2026brain4fms}. These findings motivate two aspects of our approach. First, they favour compute-efficient architectures that make scaling studies more affordable. Second, they argue for evaluations that probe what a model has learned about the data distribution itself. Generative evaluation, i.e. judging a model by the realism of the data it samples, has been used for MEG-GPT~\citep{huang2026meggpt} and in a more recent EEG foundation model~\citep{hwang2026bfm2}: this is the perspective we adopt here.

\paragraph{MEG data characteristics.} Neuronal populations exhibit transient oscillations in canonical frequency bands: delta (1--4\,Hz), theta (4--8\,Hz), alpha (8--13\,Hz), beta (13--30\,Hz), and gamma (30--45\,Hz), whose spatial distribution varies across the cortex~\citep{hari1997oscillations,brookes2011meg}. This structure carries information over both short and long timescales: fast bursts of oscillatory activity with lifetimes on the order of 100\,ms~\citep{baker2014hmm} and slow wave activity on the order of seconds. Modelling the slow dynamics motivates sequence models with a long temporal context.

As noted above, we study source reconstructed, parcellated MEG, which is calculated by projecting the sensor data onto the cortex and estimating the activity for particular regions of interest (ROIs) defined by an atlas; these regions are referred to as `parcels'. In this work, we look at 100 parcels across the cortex.

\paragraph{State-space models.} Building on structured state-space models such as S4~\citep{gu2022s4}, the Mamba architecture introduced a selective, input-dependent scan that lets the model choose what to retain or discard at each step while preserving linear-time, constant-memory recurrence~\citep{gu2023mamba}. Subsequent refinements, notably Mamba-2~\citep{dao2024mamba2} and Mamba-3~\citep{lahoti2026mamba3}, further advanced the frontier of performance and efficiency; Mamba-3 in particular introduced a complex-valued state update, which can represent damped oscillatory modes that a real-valued diagonal recurrence cannot~\citep{gu2022s4d,rusch2025linoss}. We adopt Mamba-3 as the backbone of MEG-Mamba.

SSMs have begun to appear in neural foundation models, but not, to our knowledge, as an autoregressive backbone for MEG or EEG. Mamba-based EEG foundation models such as FEMBA~\citep{tegon2025femba}, EEGMamba~\citep{wang2025eegmamba}, and EEGM2~\citep{hong2025eegm2} use \emph{bidirectional} Mamba encoders with reconstruction objectives, which yield representations but not an autoregressive generative model; CaMBRAIN~\citep{durgam2026cambrain} uses a causal Mamba for real-time EEG inference but with a self-supervised, non-generative objective. Beyond MEG and EEG, NeuroMamba~\citep{choi2025neuromamba} pre-trains a Mamba backbone autoregressively on fMRI, and SSMs have been benchmarked for autoregressive forecasting of neuronal spiking~\citep{minnick2026spikeprophecy}.

\newpage

\section{Data} \label{sec:data}

\subsection{Datasets} \label{sec:datasets}

We used the Cam-CAN dataset~\citep{shafto2014camcan,taylor2017camcan}, which consists of a large cohort of healthy adults (aged 18--89 years, mean 55.4 $\pm$ 18.3 years; 277 female). The MEG data were recorded using a 306-channel Elekta Neuromag VectorView system during eyes-closed rest, a sensorimotor task, and a passive (audio-visual) sensory task.

MEG-Mamba was pre-trained on the resting-state data only. In total, MEG-Mamba was pre-trained on 559 resting-state sessions (one session per subject), comprising 87\,h (sampled at 250\,Hz). We held out a random 10\% of the resting-state sessions (62 sessions, 9.6\,h), so that evaluation sessions came from subjects unseen during pre-training.

For task conditioning, we used the passive and sensorimotor tasks: in the passive task, participants were presented with unimodal auditory or visual stimuli; in the sensorimotor task, they responded with a button press to audio-visual stimuli. Together these yielded three event types: visual stimuli, auditory stimuli, and button presses.

\subsection{MEG Processing: Preprocessing, Source Reconstruction, and Parcellation} \label{sec:processing}

All MEG processing was performed with osl-dynamics~\citep{gohil2024osld}. We used the MaxFiltered data with movement compensation provided by Cam-CAN. We notch filtered the data at 50 and 100\,Hz, bandpass filtered (0.5--80\,Hz; fifth-order Butterworth), and downsampled to 250\,Hz. Bad segments and channels were identified and excluded automatically, separately for magnetometers and gradiometers, using generalised ESD outlier detection on signal variance (2\,s windows on both the signal and its temporal derivative). Structural MRIs were coregistered to the MEG data with RHINO, and a single-shell forward model was computed on an 8\,mm volumetric grid. Source activity was estimated using a unit-noise-gain-invariant LCMV beamformer (data rank 60, regularisation 0.05) and parcellated into the 100 ROIs in the Schaefer atlas~\citep{schaefer2018,osld_schaefer100} by taking the first principal component of the voxel time courses within each parcel; no leakage reduction (orthogonalisation) or sign flipping was applied. Sessions whose overall signal variance was an outlier (modified $z$-score $>5$ on the log standard deviation) were excluded. Finally, each parcel time course was z-scored within each session.

\section{Method} \label{sec:method}

\subsection{Model Architecture} \label{sec:architecture}

\begin{figure}[!ht]
  \centering
  \includegraphics[width=\linewidth]{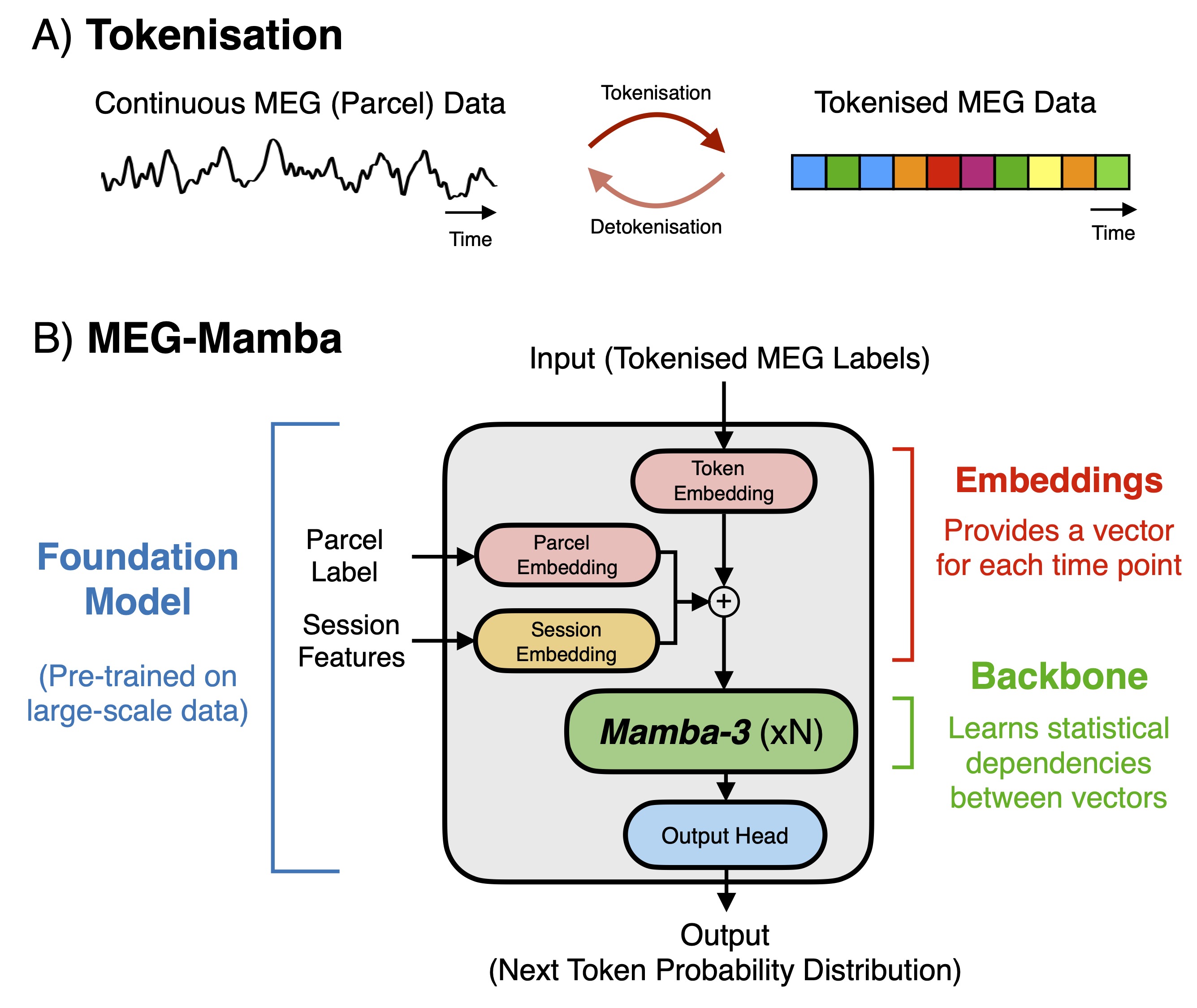}
  \caption{\textbf{Overview of MEG-Mamba.} A) Continuous source reconstructed, parcellated MEG data are mapped to a sequence of discrete tokens by a learned tokeniser model. B) MEG-Mamba is an autoregressive foundation model that predicts the next token (each parcel is modelled independently); the backbone stacks $N$=4 Mamba-3 blocks.}
  \label{fig:architecture}
\end{figure}

\Cref{fig:architecture} gives an overview of MEG-Mamba, which comprised embedding layers, a Mamba-3 backbone, and an output head; we describe each in turn below.

\paragraph{Tokenisation.} MEG-Mamba operated on discrete tokens rather than continuous values (\Cref{fig:architecture}A). We used the \emph{causal} variant of the learnable, sample-level tokeniser of \citet{cho2026tokenizer}, which discretised each parcel's signal into one of $V=92$ tokens (at each time point) and supported near-lossless detokenisation, so that data generated in token space could be rendered back as continuous parcel MEG. In the causal variant, the token at a given time point depended only on the current and past signal, so no information from future samples leaked into it.

We released a pre-trained tokeniser with MEG-Mamba. This tokeniser was trained on Schaefer100-parcellated Cam-CAN data at 250\,Hz (\Cref{sec:datasets}), pooling across resting-state, passive, and sensorimotor task recordings: a random subsample of 556 sessions from 407 participants (aged 18--89 years; 204 female), totalling 63.7\,h, stratified by task. 10\% of these data were held out for validation, and a further 92 sessions (10.6\,h) were held out as a test set, on which the tokeniser recovered 97.8\% of the variance of the original signal after detokenisation.

\paragraph{Embeddings.} At each time point, the input token was mapped to a 256-dimensional token embedding (\Cref{fig:architecture}B). To this we added two further embeddings of the same dimension: a \emph{parcel} embedding, a learned per-parcel lookup (one vector for each of the 100 parcels); and a \emph{session} embedding, produced by a small multilayer perceptron applied to precomputed session-level features. For each session and parcel, these features were the normalised token unigram histogram (92 values) and token bigram transition matrix, each square-root transformed and standardised, with the bigram block reduced to 200 dimensions by principal component analysis (PCA). The standardisation and PCA were fitted on training sessions only and frozen, so unseen sessions were processed identically. The resulting 292-dimensional feature vector was mapped through two 128-unit hidden layers. The three embeddings were summed to form a single input vector per time point. Because the session features were computed directly from the recording, a session embedding could be obtained for new recordings without any further training, facilitating zero-shot inference.

\paragraph{Backbone.} The resulting sequence of input vectors was processed by a backbone of 4 stacked Mamba-3 blocks~\citep{lahoti2026mamba3} of width 256, which modelled temporal dependencies through the selective scan described in \Cref{sec:related}. Each block applied a pre-norm residual update with a multi-input multi-output (MIMO) Mamba-3 mixer (state size 64, expansion factor 2, head dimension 64, MIMO rank 4) followed by a SwiGLU feed-forward layer, using RMSNorm throughout.

\paragraph{Output head.} A final RMSNorm followed by a linear layer mapped the backbone output at each time point to logits over the $V=92$ tokens, which a softmax turned into a probability distribution over the next token.

\subsection{Pre-training Objective} \label{sec:pretraining}

MEG-Mamba was pre-trained by autoregressive next-token prediction. Given the tokens up to time $t$ for a parcel, together with the parcel and session embeddings, the model predicted the token at time $t{+}1$, and was trained to minimise the cross entropy between the predicted distribution and the true next token. To avoid penalising predictions made before the model had accumulated sufficient context, the first $W=50$ time points of each sequence were excluded from the loss and served as warm-up. Writing $x^{(s,j)}_t$ for the token of parcel $j$ in session $s$ at time $t$, and $T=1{,}000$ for the context length, the pre-training loss was
\begin{equation}
  \mathcal{L}(\theta) = -\,\mathbb{E}_{(s,j)} \left[ \frac{1}{T-W} \sum_{t=W}^{T-1} \log p_\theta\!\left( x^{(s,j)}_{t+1} \,\middle|\, x^{(s,j)}_{1:t},\, j,\, s \right) \right],
  \label{eq:loss}
\end{equation}
where $\theta$ denotes the model parameters, conditioning on $j$ and $s$ was through the parcel and session embeddings, and the expectation was taken over training windows drawn from all parcels and sessions.

\subsection{Task Conditioning} \label{sec:conditioning}

MEG-Mamba was pre-trained on eyes-closed resting-state data (\Cref{sec:datasets}), so the pre-trained model generated resting-state-like data. To generate brain activity in response to particular tasks, referred to as \emph{task-evoked responses}, we conditioned the model on the stimulus while keeping the pre-trained backbone frozen.

We added a fourth term alongside the token, parcel, and session embeddings: a \emph{stimulus} embedding driven by a multi-hot boxcar over three event categories: visual, auditory, and motor (button press). Each event switched its category on for $1$\,s from its onset; at each time point the active categories selected and summed the corresponding embedding vectors, which were added to the summed input embedding $h_t$ ($h_t \leftarrow h_t + u_t W_{\mathrm{stim}}$, with $u_t\in\{0,1\}^{3}$ and $W_{\mathrm{stim}}\in\mathbb{R}^{3\times256}$). Co-occurring events combined additively: a bimodal audio-visual stimulus activated both the visual and auditory embeddings, and an overlapping stimulus and response summed. The model was therefore told only when, and which type of, event occurred, never the response itself.

To allow the stimulus to modulate neuronal dynamics, we additionally applied low-rank adaptation (LoRA)~\citep{hu2022lora} with rank $16$ and scaling $\alpha=32$, injecting adapters into the input and output projections of each Mamba-3 block; because the stimulus entered at the input, the adapted input projections acted on stimulus-bearing activations. The stimulus embedding and the LoRA up-projections were zero-initialised, so the conditioned model was identical to the pre-trained one at the start of fine-tuning. All other parameters, including the embeddings, backbone, final norm, and output head, were frozen, so only the stimulus embedding and the LoRA parameters were learned (${\approx}167$k parameters, about $5\%$ of the pre-trained model). Fine-tuning used the same next-token objective as pre-training (\Cref{eq:loss}).

\section{Experiments} \label{sec:experiments}

To evaluate MEG-Mamba, we examined how realistic the spatial, spectral, and temporal properties of the data it generated were. Additionally, we examined whether the model could be steered to produce realistic task-evoked responses.

\subsection{Experimental Setup} \label{sec:setup}

\paragraph{Pre-training.} MEG-Mamba was pre-trained on the 559 training sessions (\Cref{sec:datasets}), used in full and divided into non-overlapping 4\,s windows that were modelled one parcel at a time. This gave 7.8M single-parcel sequences per epoch, or ${\sim}15$k optimisation steps with batches of 512. We trained for three epochs with AdamW ($\beta=(0.9,0.95)$, weight decay 0.01) at a learning rate of $10^{-4}$, reached by a linear warm-up over the first 2{,}000 steps and then held constant. Gradients were clipped to a norm of $1$, and training used bfloat16 mixed precision. \Cref{tab:config} gives the full configuration. We report a single training run (seed 42). Held-out cross entropy decreased at every epoch, so the checkpoint with the lowest held-out cross entropy was also the final one. The pre-trained weights, tokeniser, full configuration, and training code are publicly available.\footnote{\url{https://github.com/OHBA-analysis/MEG-Mamba}.}

\paragraph{Evaluation metrics.} We report next-token cross entropy (nats per token, as in \Cref{eq:loss}) and top-1 accuracy over 4\,s windows drawn uniformly from the 62 held-out sessions; the train reference is the identically sized sample drawn over all 559 training sessions (300 batches of $512$ single-parcel sequences, i.e. 153{,}600 sequences, 1{,}536 windows $\times\,100$ parcels, each). Training curves (\Cref{fig:training}A, left and middle) show running averages over each epoch; the generalisation gap (\Cref{fig:training}A, right) is the held-out minus the training value at the final epoch, using the training sample above. With 92 token IDs, chance-level top-1 accuracy is $1/92 \approx 1.1\%$.

To compare generated and real data, we computed power spectra for each parcel with Welch's method (2\,s Hann windows, 50\% overlap) and integrated them over the canonical frequency bands (\Cref{sec:related}). The band-power correlation $r$ is the Pearson correlation across the 100 parcels between real and generated band power (each first averaged over the 62 held-out sessions). Relative power maps show, for each band and parcel, that band's power relative to the bandwidth-weighted mean power across the five bands for the same parcel, on session-averaged spectra.

Time-frequency representations of the task-evoked response used Morlet wavelets over 6--30\,Hz in 0.5\,Hz steps with a fixed 4 cycles per frequency, decimated by 3, on epochs of $-0.5$ to $1.5$\,s for all event types (computed on epochs padded by 0.5\,s on each side and cropped after the transform, to avoid edge effects), log-ratio baseline-corrected against [$-0.5$, $-0.1$]\,s and averaged over trials.

For the embedding analysis (\Cref{fig:embeddings}), we applied PCA to the 100 learned parcel embeddings and, separately, to the session embeddings of all 621 sessions, where each session's 100 per-parcel embeddings were averaged (mean over parcels) to give one 256-d vector per session.

\begin{table}[!ht]
  \caption{MEG-Mamba configuration.}
  \label{tab:config}
  \centering
  \small
  \resizebox{\linewidth}{!}{%
  \begin{tabular}{@{}llll@{}}
    \toprule
    \multicolumn{2}{@{}l}{\textbf{Architecture}} & \multicolumn{2}{l@{}}{\textbf{Pre-training}} \\
    \midrule
    Backbone            & $4\times$ Mamba-3 blocks        & Objective        & Next-token cross entropy \\
    Model width         & $256$                           & Context length   & $1{,}000$ samples ($4$\,s) \\
    SSM state size      & $64$                            & Loss warm-up     & first $50$ samples excluded \\
    Expansion factor    & $2$                             & Batch size       & $512$ sequences \\
    Head dimension      & $64$                            & Optimiser        & AdamW, $\beta=(0.9,0.95)$ \\
    MIMO rank           & $4$                             & Learning rate    & $10^{-4}$ ($2$k-step warm-up, then constant) \\
    Feed-forward        & SwiGLU                          & Weight decay     & $0.01$ \\
    Normalisation       & RMSNorm (pre-norm)              & Epochs           & $3$ \\
    Vocabulary          & $92$ tokens (causal tokeniser)  & Precision        & bfloat16 \\
    Embeddings          & token + parcel + session ($256$-d) & Hardware      & $1\times$ NVIDIA A100 ($80$\,GB) \\
    Parameters          & $3.4$M                          & Training cost    & $22$ GPU-hours \\
    \midrule
    \multicolumn{2}{@{}l}{\textbf{Data}} & \multicolumn{2}{l@{}}{\textbf{Task conditioning}} \\
    \midrule
    Dataset             & Cam-CAN, eyes-closed rest       & Trainable        & stimulus emb.\ + LoRA ($167$k) \\
    Sessions            & $559$ train / $62$ held out     & LoRA             & rank $16$, $\alpha=32$ \\
    Pre-training data   & ${\approx}87$\,h                & Adapted layers   & Mamba input/output projections \\
    Parcellation        & Schaefer $100$ ROIs             & Learning rate    & $10^{-3}$, cosine decay, $10$ epochs \\
    Sampling rate       & $250$\,Hz                       & Events           & visual, auditory, button press ($1$\,s boxcar) \\
    \bottomrule
  \end{tabular}}
\end{table}

\paragraph{Resting-state generation.}
For each held-out session, we took a random window of the recording, prompted the model with its first 4\,s and generated a further 30\,s autoregressively, conditioned on the session's own parcel and session embeddings. Tokens were drawn from the full predicted distribution (temperature 1, without top-$k$ or nucleus truncation). The generated tokens were detokenised and compared with the real recording of the same session using the metrics above.

\paragraph{Task conditioning.} We fine-tuned on the continuous passive and sensorimotor recordings of 530 subjects whose resting-state data were used in pre-training (852 sessions: 522 passive, 330 sensorimotor; 69.4\,h); a further 28 such subjects were kept aside for validation. Sensorimotor sessions in which button presses were not recorded (225 of 617) were excluded. We modelled three parcels (primary visual, auditory, and motor cortex), chosen as the Schaefer parcels nearest to canonical coordinates; each parcel received the full multi-hot stimulus timeline.

Session embeddings for the task recordings were computed from each recording with the frozen feature transform (\Cref{sec:architecture}). Training sampled random 8.2\,s windows (2{,}048 time points) of the continuous recordings, so that, as during generation, each window contained several events with long preceding context.

We trained with AdamW ($\beta=(0.9,0.95)$, no weight decay) at a peak learning rate of $10^{-3}$ (200-step warm-up, then cosine decay to zero), with batches of 32 windows, gradient clipping at 1, and bfloat16 precision, for 10 epochs (28{,}600 steps; 4.4\,h on a single NVIDIA L4), and used the final checkpoint.

To generate task-evoked data (\Cref{fig:tfr}), we used the 62 subjects held out of pre-training (105 sessions: 62 passive, 43 sensorimotor), none of whom were seen during fine-tuning. For each session, we prompted the model with the first 5\,s of the real recording and generated the remainder of the session autoregressively (2.5--9\,min), drawing tokens from the full predicted distribution and switching the stimulus term on at the times of that session's real events. Real and generated data were both passed through the tokeniser round trip, epoched around the same events, and compared through trial-averaged time-frequency responses (8{,}513 auditory, 8{,}472 visual, and 5{,}241 button-press trials; one generation per session).

\subsection{Results} \label{sec:results}

\begin{figure}[!ht]
  \centering
  \includegraphics[width=\linewidth]{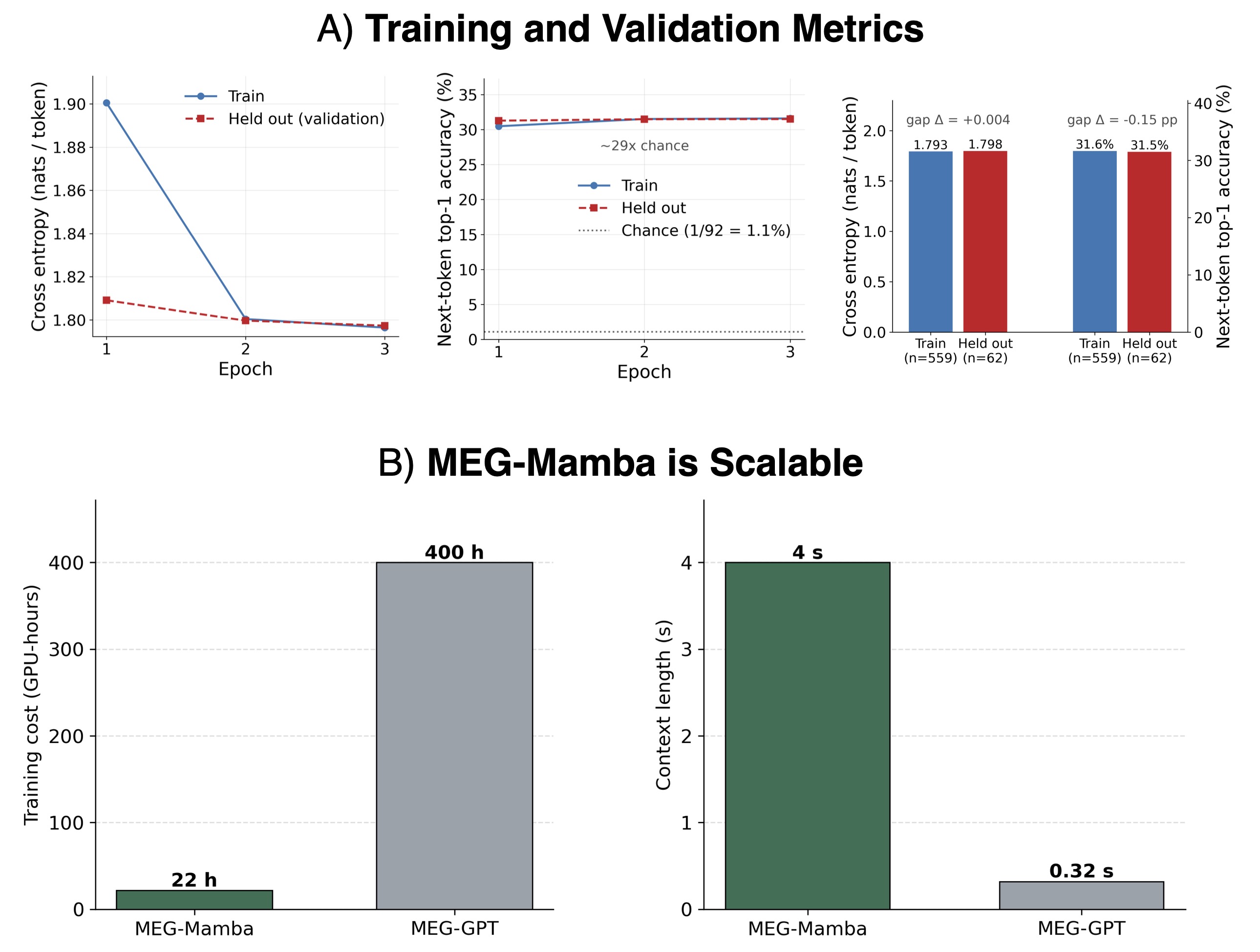}
  \caption{\textbf{Training dynamics and scalability.} A) Cross entropy (left) and next-token top-1 accuracy (middle) on the training and held-out sets over epochs. The small generalisation gap (right) indicates negligible overfitting. B) Pre-training cost (left) and context length (right) of MEG-Mamba and MEG-GPT.}
  \label{fig:training}
\end{figure}

\paragraph{Efficient training.} Training and held-out cross entropy decreased together over epochs and converged to nearly identical values, leaving a negligible generalisation gap (\Cref{fig:training}A). The model reached a held-out cross entropy of 1.80 nats per token and 31.5\% next-token top-1 accuracy, roughly $29\times$ chance, indicating that it learned genuinely predictive structure rather than memorising the training set. This was achieved efficiently: MEG-Mamba pre-trained in 22 GPU-hours, against 400 for MEG-GPT, while modelling a 4\,s context rather than 0.32\,s (\Cref{fig:training}B), an order-of-magnitude improvement in both training cost and context length on the same dataset.

\begin{figure}[!ht]
  \centering
  \includegraphics[width=\linewidth]{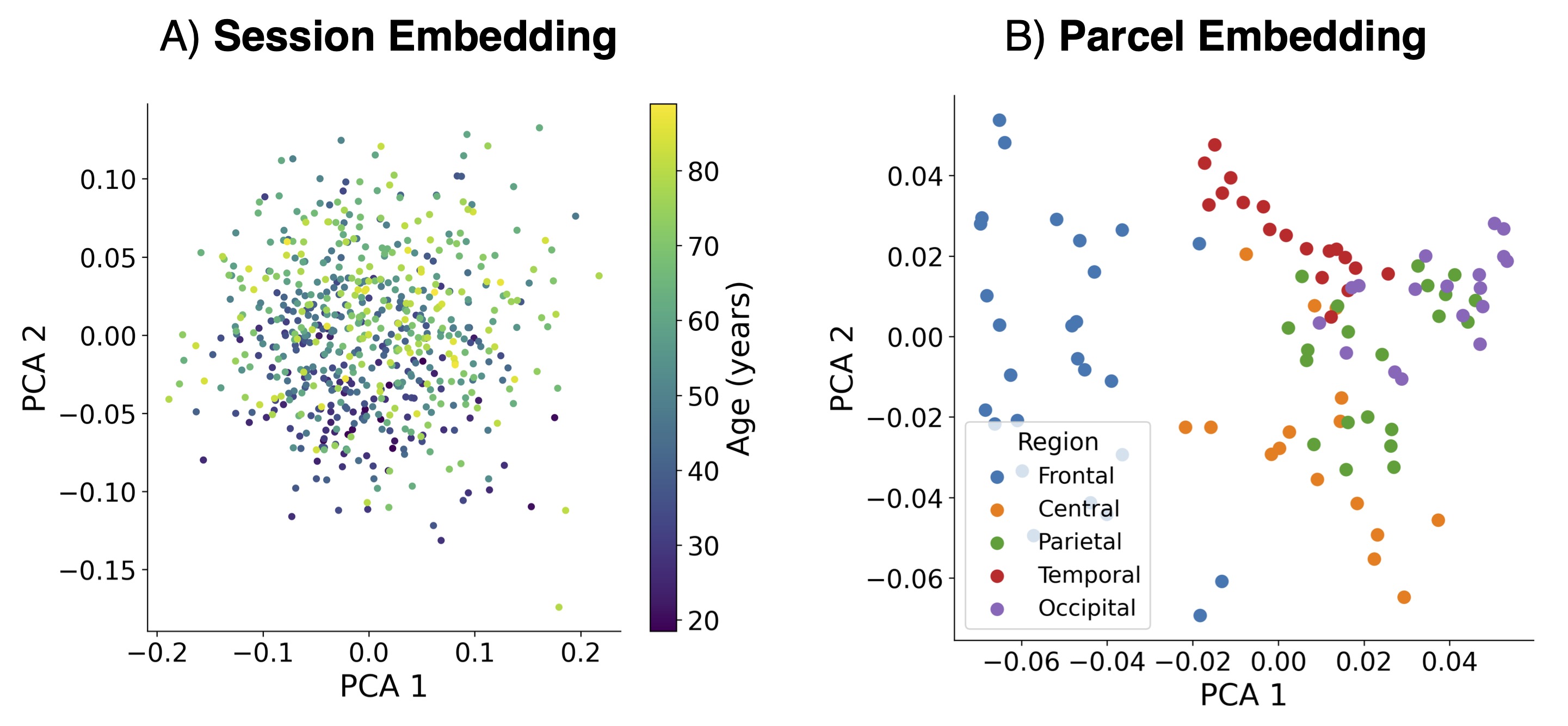}
  \caption{\textbf{MEG-Mamba learns meaningful embeddings.} PCA projections of the learned embeddings. A) Session embeddings encode participant age. B) Parcel embeddings recover the spatial organisation of the cortex: parcels cluster by cortical region (frontal, central, parietal, temporal, occipital).}
  \label{fig:embeddings}
\end{figure}

\begin{figure}[!ht]
  \centering
  \includegraphics[width=\linewidth]{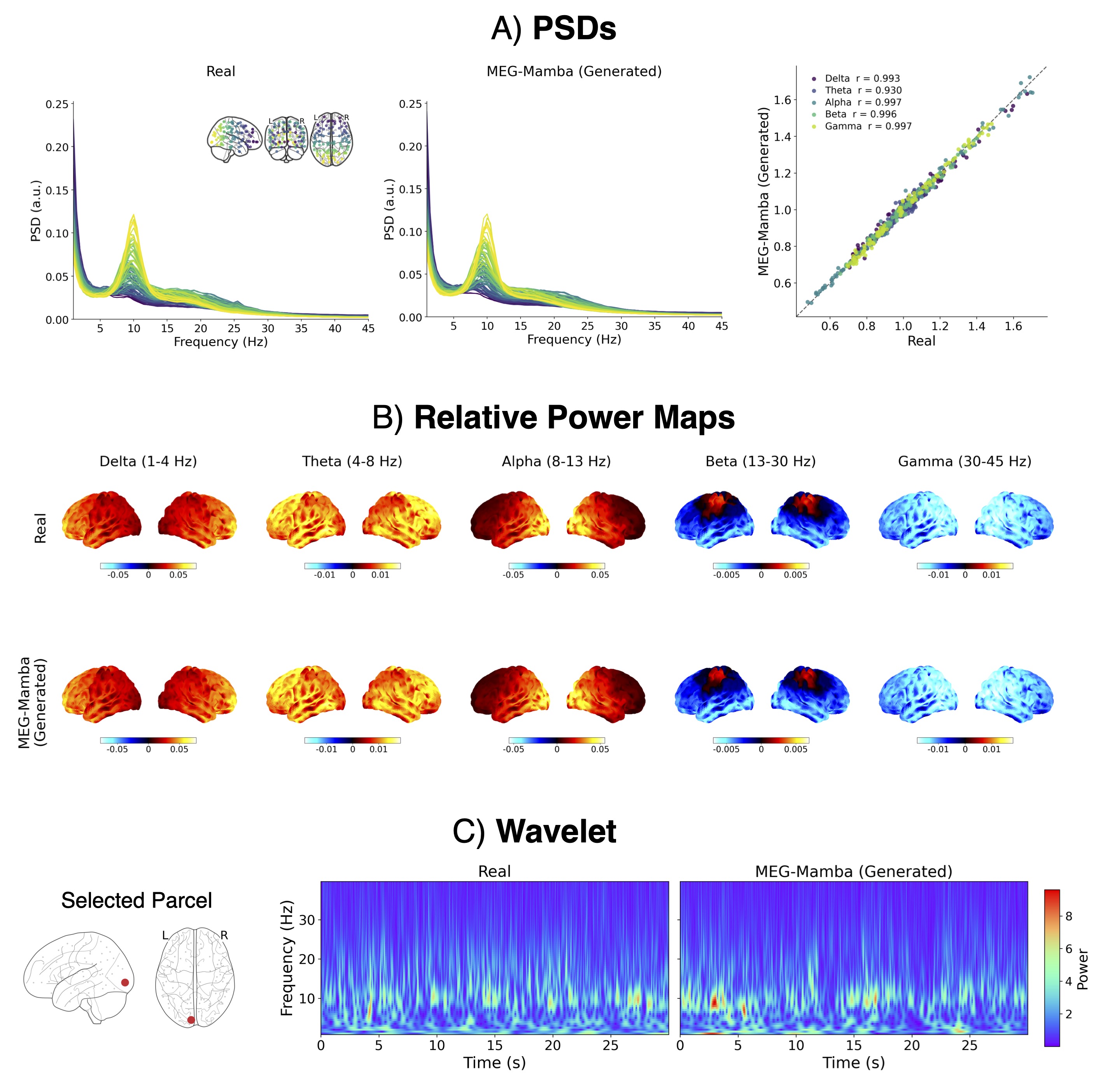}
  \caption{\textbf{MEG-Mamba produces realistic, high-fidelity synthetic MEG data.} Data are generated for held-out subjects; their spatio-spectral characteristics (A, B) and dynamics (C) closely match real data. A) Power spectral densities of each parcel for real (left) and generated (middle) data, and real versus generated band power for each parcel (right). B) Relative power maps for each canonical band. C) Wavelet time-frequency representation of $30$\,s of real and generated data at an occipital parcel.}
  \label{fig:generated-data}
\end{figure}

\begin{figure}[!ht]
  \centering
  \includegraphics[width=0.85\linewidth]{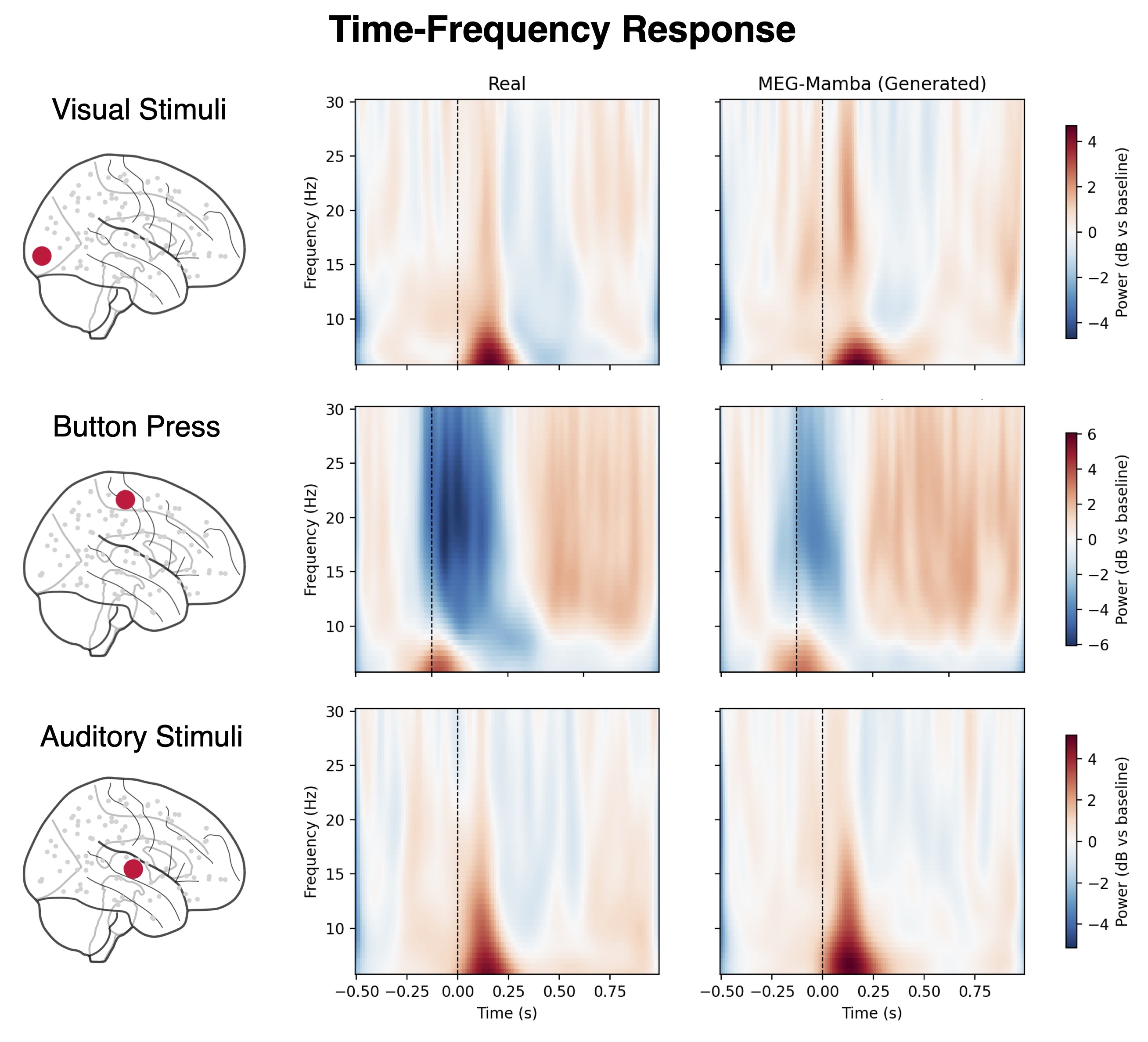}
  \caption{\textbf{MEG-Mamba reproduces task-evoked responses for held-out subjects.} Trial-averaged time-frequency responses (log-ratio baseline, dB; shown from $-0.5$ to $1$\,s) of real data (left) and data generated by the stimulus-conditioned model (\Cref{sec:conditioning}; right), for visual stimuli (top; left primary visual cortex), button presses (middle; left primary motor cortex), and auditory stimuli (bottom; right primary auditory cortex), in the $62$ subjects held out of pre-training and fine-tuning. Generated data recover the characteristic evoked spectral patterns.}
  \label{fig:tfr}
\end{figure}

\paragraph{MEG-Mamba learns meaningful embeddings.} A useful foundation model should organise its internal representations around the structure of the data. Examining the learned embeddings via PCA (\Cref{fig:embeddings}), we found that the parcel embeddings recovered the spatial organisation of the cortex: with no explicit spatial supervision, parcels clustered by cortical region (frontal, central, parietal, temporal, and occipital; \Cref{fig:embeddings}B). The session embeddings, in turn, captured subject-level structure, varying with participant age (\Cref{fig:embeddings}A). MEG-Mamba's learned embeddings thus encode interpretable anatomical and demographic information.

\paragraph{High-fidelity generation for unseen subjects.} We next assessed the realism of data generated for held-out subjects (\Cref{fig:generated-data}). Generated and real (resting-state) data had closely matching power spectral densities, with real-versus-generated band-power correlations of $r\geq0.93$ across all canonical bands (\Cref{fig:generated-data}A). The spatial distribution of band power was also faithfully reproduced, with relative power maps from delta to gamma matching those of real data (\Cref{fig:generated-data}B). Finally, wavelet time-frequency representations showed that the generated data reproduced the transient, non-stationary dynamics of real MEG (\Cref{fig:generated-data}C). MEG-Mamba thus captures the general spatial, spectral, and temporal structure of MEG across a large population.

\paragraph{Conditional generation of task-evoked responses.} Beyond resting-state generation, MEG-Mamba can be steered to produce task-specific data. Sampling from the stimulus-conditioned model (\Cref{sec:conditioning}) yielded time-frequency responses that closely matched those of real task data at the task-relevant parcels (\Cref{fig:tfr}). These responses were recovered for subjects unseen in both pre-training and fine-tuning, showing that the pre-trained model can be cheaply adapted to generate realistic, condition-specific neural activity.

\section{Discussion} \label{sec:discussion}

MEG-Mamba showed that a state-space backbone can surpass the generative fidelity of a transformer-based MEG foundation model (MEG-GPT) at an order of magnitude lower training cost and with a substantially longer context window. The architecture also benefits generation: a recurrent model carries a fixed-size state, so each generated time point costs the same regardless of how much context precedes it, whereas an attention-based model must attend over a growing cache. For MEG, whose dynamics combine slow fluctuations with fast transient events, the ability to model long contexts cheaply is especially valuable, and the recurrent formulation also opens the door to real-time inference.

MEG-Mamba's state-space backbone also connects it to a long line of work that models MEG dynamics with state-space models. Classical approaches such as hidden Markov models (and newer variants) infer a latent state that evolves over time, segmenting brain activity into recurring, interpretable networks~\citep{baker2014hmm,vidaurre2018hmm,gohil2022dynemo,cho2026dyneste} and underpinning much of the current state of the art in characterising fast dynamics in electrophysiological brain data. MEG-Mamba carries an explicit latent state through its selective recurrence, raising the possibility of interpreting that state in similar terms, for example by relating its evolution to known transient network states. We see this as a promising direction for future work, bridging high-performing sequence models with the interpretable dynamical descriptions valued in neuroscience.

Our generative evaluation (\Cref{sec:results}) mirrored how progress in image and audio generation is assessed: a model that reproduces the spatial, spectral, and temporal structure of held-out MEG has captured properties that a narrow downstream benchmark might miss. We view it as complementary to, not a replacement for, downstream task evaluations.

Beyond generation, MEG-Mamba's internal structure was interpretable and reusable. Its embeddings captured cortical anatomy and participant age without direct supervision (\Cref{fig:embeddings}), suggesting that these representations could transfer to downstream tasks or population-level analyses. The model was also controllable: a lightweight adapter was enough to generate realistic task-evoked responses for unseen subjects, pointing to uses in data augmentation, simulation, and in-silico experimentation, and as a generative prior for neural decoding.

\paragraph{Limitations.} MEG-Mamba was trained on a single dataset recorded at one site (Cam-CAN), and it was pre-trained on eyes-closed rest only; both the corpus ($\approx 87$\,h) and the model ($3.4$M parameters) are modest by the standards of foundation models in other domains. We assessed generative fidelity with summary statistics (spectra, power maps, and time-frequency responses) rather than a formal distributional metric, and we compared against a single prior model, MEG-GPT. Quantifying how MEG-Mamba scales with more data and parameters, and measuring generation quality more rigorously, are important next steps. A complementary and equally important next step is to benchmark MEG-Mamba on a range of downstream tasks, such as neural decoding, age regression, and clinical prediction, where its learned representations can be probed or fine-tuned and compared against existing models. EEG remains valuable and abundant, and jointly modelling EEG and MEG within a single foundation model is a further promising direction.

\newpage

\section{Conclusion} \label{sec:conclusion}

We introduced MEG-Mamba, a scalable state-space foundation model for MEG. MEG-Mamba surpassed the generative fidelity of MEG-GPT at an order of magnitude lower training cost and with a longer context window. MEG-Mamba learned meaningful, interpretable embeddings and generated high-fidelity MEG for unseen subjects. We see state-space models as a promising foundation for large-scale models of brain activity.

\bibliographystyle{unsrtnat_links}
\bibliography{references}

\end{document}